\documentclass[prd,twocolumn,showpacs]{revtex4}
\usepackage{latexsym}
\usepackage{amsmath}
\usepackage{amssymb}
\usepackage[dvips]{graphicx}
\begin{document}
%
%
%
\preprint{LAUR LA-UR-02-4736}
\title{Dynamics of broken symmetry $\lambda \, \phi^4$ field theory}
\author{Fred Cooper}
\email{fcooper@lanl.gov} \affiliation{Theoretical Division,
   Los Alamos National Laboratory, Los Alamos, NM 87545}
\author{John F. Dawson}
\email{john.dawson@unh.edu}
\affiliation{Department of Physics,
   University of New Hampshire, Durham, NH 03824}
\author{Bogdan Mihaila}
\email{bogdan.mihaila@unh.edu}
\affiliation{Physics Division,
   Argonne National Laboratory, Argonne, IL 60439}

\date{\today}
\begin{abstract}
We study the domain of validity of a Schwinger-Dyson approach to
non-equilibrium dynamics when there is explicit broken symmetry.  We
perform exact numerical simulations of the one- and two-point
functions of a single component $\lambda \, \phi^4$ field theory in
1+1 dimensions in the classical field theory case, for initial
conditions where $\langle \phi(x) \rangle \neq 0$.  We compare these
results to two self-consistent truncations of the SD equations which
ignore three-point vertex function corrections.  The first
approximation, which we called the bare vertex approximation, sets the
three-point function to one.  It gives a promising description for
$\langle \phi(x) \rangle = \phi(t)$.  The second approximation, which
ignores higher in $1/N$ corrections to the 2PI generating functional
is not as accurate for $\phi(t)$ for the case $N=1$.  Both
approximations have serious deficiencies in describing the two-point
function when $\phi(0) \gtrsim 0.4$.

\end{abstract}
\maketitle
%
%
%
\section{Introduction}
\label{s:introduction}

Recently there has been much effort in finding approximation schemes
to study the dynamics of phase transitions that go beyond a mean field
theory (leading order in large-$N$) approach. This is an important
endeavor if one wants a first principles understanding of the dynamics
of quantum phase transitions.  In a previous set of papers, we have
introduced a composite field $\chi$ and studied the validity of a
next-to-leading order (NLO) truncation scheme of the Schwinger-Dyson
(SD) hierarchy of equations, in quantum mechanics~\cite{r:MCD01} as
well as in 1+1 dimensional field theory~\cite{r:BCDM01}.  We called
this energy-conserving approximation, the bare vertex approximation
(BVA).  A parallel set of investigations by Berges
et.~al.~\cite{r:BC01,r:AB01,r:AB02,r:B02,r:AABBS}, have looked at a
related NLO approximation based on the two-particle irreducible
formalism (2PI-1/N), which is also energy-conserving.  These
investigators have pointed out that when there is explicitly broken
symmetry, the BVA contains terms not included in the $1/N$ resummation
at next-to-leading order~\cite{r:AABBS}.

In this rapid communication, we present the first numerical
simulations of the dynamics for a single field 1+1 dimensional
$\phi^4$ field theory for explicit symmetry breaking, where a new
scale is set by initial values of the field.  We compare the BVA with
the 2PI-1/N for the classical field theory case, where exact
comparison with numerical simulations are possible.  The differences
between the classical and quantum field theory, are discussed in
detail in Refs.~\cite{r:BCDM01,r:CKR}, and the unbroken symmetry
version of this paper is described in detail in
Ref.~\cite{r:BCDM01}. Here, we focus on the difference between these
two approximations when compared with exact simulations.  We also
present a new algorithm for solving the SD equations for the
broken-symmetry problem.

The long-term goal of this work is finding approximation schemes which
are accurate at the small values of N relevant to the case of
realistic quantum field theories.  In order to maximize the possible
differences between approximation schemes and the exact result, we
choose to study the O(N) model for N=1.  Based on our previous studies
of the quantum mechanical version of this model~\cite{r:MCD01} we
expect that by increasing~N the differences will diminish.  This is
due to the fact that the SD formalism is related, but not identical,
with approximations based on the large N expansion.  As such, terms
which may constitute next-to-next-to-leading order (NNLO) for the
2PI-1/N expansion are in fact NLO in the BVA.

In our previous work~\cite{r:BCDM01} we have shown that for the
symmetric case, the BVA and 2PI-1/N gave results of similar quality.
However, the agreement may have been misleading as in that study we
only dealt with the restrictive case of zero initial field expectation
values. That choice was made solely for the purpose of reducing the
computational cost of the calculation.  In the symmetric case, the
restrictions imposed by the choice of initial conditions,
significantly simplifies the equations we need to solve, since many
terms vanish identically by construction. In this paper we report the
first results of the full calculation.


Since the 2PI-1/N expansion is a subset of the BVA, we can use the
same code in order to generate predictions for both approximation
schemes. The differences between the BVA and 2PI-1/N, and between
these approximations and the exact result are non-trivial. We will
show that the BVA gives a better description of the order parameter
$\phi(t)$ than the 2PI-1/N. However both approximations suffer from
deficiencies when describing the equilibration time of the two-point
function when the initial value of the order parameter $\phi(0)$
exceeds a certain fraction of the mass parameter $\mu$.  Having
established the efficacy of the BVA for describing the order parameter
$\langle \phi(x,t) \rangle$, we intend to apply this method to the
$O(4)$ linear sigma model in 3+1 dimensions to study the dynamics of
chiral symmetry breaking in a realistic effective field theory.

%
%
\section{Equations of the 2PI effective action approach}
\label{s:equs2PI}

The truncation of the SD equations we are proposing can be obtained
from a 2PI effective action~\cite{r:CJT,r:LW,r:Baym62}. Other
approaches leading to these equations are found
in~\cite{r:MCD01,r:AABBS}. Using the extended fields notation, $
\phi_\alpha(x) = [ \chi(x), \phi_1(x), \phi_2(x), \ldots , \phi_N(x)]
\>, $ the effective action can be written as:
\begin{equation*}
   \Gamma[\phi_\alpha,G]
   =
   S_{\text{cl}}[\phi_\alpha]
   +
   \frac{i}{2} {\rm Tr} \ln G^{-1}
   +
   \frac{i}{2} {\rm Tr} G_0^{-1} G
   +
   \Gamma_2[G] \>,
\end{equation*}
where $ \Gamma_2[G]$ is the generating functional of the 2-PI
graphs,
and the classical action
\begin{multline}
   S_{\text{cl}}[\phi_\alpha]
   =
   \int dx \,
   \Bigl \{ \,
   - \frac{1}{2}
   \phi_i(x) \,
   \bigl [ \, \Box + \chi(x) \, \bigr ] \, \phi_i(x)
   \\
   +
   \frac{\chi^2(x)}{2 g}
   -
   \frac{\mu^2}{g} \chi(x) \,
   \Bigr \} \>,
\end{multline}
is that of $\lambda \phi^4$ field theory written in terms of the
auxiliary field $\chi$ appropriate for studying $1/N$
expansions~\cite{r:CJP}.  The integrals and delta functions
$\delta_{\mathcal{C}}(x,x')$ are defined on the closed time path
(CTP) contour, which incorporates the initial value boundary
condition~\cite{r:Schwinger,r:Keldish,r:MahanI,r:MahanII}.  Here
$g=\lambda/N$, and $\lambda$ is held fixed if we are studying the
large-$N$ limit. The approximations we are studying include only
the two-loop contributions to $\Gamma_{2}$.

\begin{figure}[h]
   \centering
   \includegraphics[width=2.in]{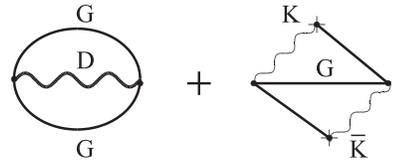}
   \caption{
            Graphs included in the 2PI effective action $\Gamma_2$.}
   \label{f:fig1}
\end{figure}

The Green function $G^{-1}_{0\,\alpha \beta}[x](x,x')$ is defined as
follows:
\begin{align}
   G_{0\,\alpha \beta}^{-1}[\phi](x,x')
   &=
   - \frac{ \delta^2 S_N[\phi;\mathrm{j}_\phi] }
          {\delta \phi_{\alpha}(x) \, \delta \phi_{\beta}(x') }
   \label{SD.eq:ginvdef} \\
   &=
   \begin{pmatrix}
      D^{-1}_0(x,x')   & \bar{K}_{0\,j}^{-1}(x,x') \\
      K_{0\,i}^{-1}(x,x') & G_{0\,i j}^{-1}(x,x')
   \end{pmatrix}  \>,
   \notag
\end{align}
where
\begin{align}
   D_0^{-1}(x,x')
   &=
      - \frac{1}{g}
    \, \delta_{\mathcal{C}}(x,x') \>,
   \notag \\
   G_{0\,ij}^{-1}[\chi](x,x')
   &=
   \bigl [
      \Box
      + \chi(x)
   \bigr ] \, \delta_{ij} \delta_{\mathcal{C}}(x,x') \>,
   \label{SD.eq:dginv}
\end{align}
and the off-diagonal elements are
\begin{equation*}
   K_{0\,i}^{-1}[\phi](x,x')
   =
   \bar{K}_{0\,i}^{-1}[\phi](x,x')
   =
   \phi_i(x) \, \delta_{\mathcal{C}}(x,x') \>.
\end{equation*}
The exact Green function $G_{\alpha \beta}[j](x,x')$ is defined by:
\begin{align}
   G_{\alpha \beta}[j](x,x')
   &=
   \frac{\delta^2 W[j]}{\delta j_\alpha(x) \, \delta j_\beta(x')}
   \label{SD.eq:GGdef} \\
   &=
   \begin{pmatrix}
      D(x,x')     & K_j(x,x') \\
      \bar{K}_i(x,x') & G_{i j}(x,x')
   \end{pmatrix}  \>,
   \notag
\end{align}
The exact equations following from the effective action are:
\begin{gather}
   \bigl [
      \Box
      + \chi(x)
   \bigr ] \, \phi_i(x)
   +
   K_i(x,x) / i
   = 0 \>,
   \label{e:GammachieqBVA} \\
   \chi(x)
   = \mu^2+
   \frac{g}{2}
      \sum_i
      \bigl [
         \phi_i^2(x)
         +
         G_{ii}(x,x)/i
      \bigr ] \>,
   \notag
\end{gather}
and
\begin{equation}
   G_{\alpha \beta}^{-1}(x,x')
   =
   G_{0\, \alpha \beta}^{-1}(x,x')
   +
   \Sigma_{\alpha \beta}(x,x')  \>,
   \label{e:GGinvGinvSigma}
\end{equation}
where
\begin{align}
   \Sigma_{\alpha \beta}(x,x')
   &=
   \begin{pmatrix}
      \Pi(x,x')          & \Omega_j(x,x') \\
      \bar\Omega_i(x,x') & \Sigma_{ij}(x,x')
   \end{pmatrix}
   \notag \\
   &=
   \frac{2}{i} \,
   \frac{\delta \Gamma_2[G]}{\delta G_{\alpha\beta}(x,x')} \>.
   \label{e:Sigmasdefs}
\end{align}
In the BVA, we keep in $\Gamma_2[G]$ only the graphs shown in
Fig.~\ref{f:fig1}, which is explicitly
\begin{align}
   \Gamma_2[G]
   &=
   - \frac{1}{4}
   \iint dx \, dy \,
   \bigl [ \,
      G_{ij}(x,y) \, G_{ji}(y,x) \, D(x,y)
   \notag \\ & \qquad
      +
      2 \, \bar{K}_i(x,y) \, K_j(x,y) \, G_{ij}(x,y) \,
   \bigr ] \>.
   \label{e:Gamma2}
\end{align}
The self-energy, given in Eq.~\eqref{e:Sigmasdefs}, then reduces
to
:
\begin{align}
   \Pi(x,x')
   &=
      \frac{i}{2}
      G_{mn}(x,x') \, G_{mn}(x,x')
   \>,
   \label{e:SigmasBVA} \\
   \Omega_i(x,x')
   &=
      i \,
      \bar{K}_m(x,x') \, G_{m i}(x,x')
   \>,
   \notag \\
   \bar\Omega_i(x,x')
   &=
      i \,
      G_{i m}(x,x') \, K_m(x,x')
   \>,
   \notag \\
   \Sigma_{ij}(x,x')
   &=
      i \,
      \bigl [
         G_{i j}(x,x') \, D(x,x')
          +
         \bar{K}_{i }(x,x') \, K_{j}(x,x')
      \bigr ]
   \,.
   \notag
\end{align}
Throughout this paper, we use the Einstein summation convention
for repeated indices.

As discussed in detail in Ref.~\cite{r:AABBS}, the second graph in
Fig.~\ref{f:fig1} is proportional to $1/N^2$ and is ignored in the
2PI-1/N approximation.  Since this term is not present when
symmetry is preserved, a comparison of these two approaches
requires studying the symmetry breaking case, which is done for
the first time here.

%
%
\section{Update equations for the Green functions}

We notice from the definitions of the matrices representing
$G_{\alpha\beta}(x,x')$ and $G_{\alpha\beta}^{-1}(x,x')$, that the
matrix elements are not inverses of one another, but instead
satisfy schematically:
\begin{gather}
   D^{-1} D
    + \bar{K}_k^{-1} \bar{K}_k
   = \delta_{\mathcal{C}}
   \>,
   \label{e:GinvG} \\
   D^{-1} K_j
    + \bar{K}_k^{-1} G_{kj} = 0
   \>,
   \notag \\
   K_i^{-1} D
    + G_{ik}^{-1} \bar{K}_k = 0
   \>,
   \notag \\
   K_i^{-1} K_j
    + G_{ik}^{-1} G_{kj}
   =
   \delta_{ij} \delta_{\mathcal{C}} \>.
   \notag
\end{gather}
Inverting Eq.~\eqref{e:GGinvGinvSigma}, we find:
\begin{align}
   D(x,x')
     = \ &
     - g \, \delta_{\mathcal{C}}(x,x')
   \\ \notag &
       + g \int_{\mathcal{C}} \mathrm{d}x_1 \,
                \Pi'(x,x_1) \ D(x_1,x')
   \>,
   \\
   G_{ij}(x,x')
   = \ &
   G_{0\,ij}(x,x') \delta_{ij}
   - \int_{\mathcal{C}} \mathrm{d}x_1 \,
   G_{0\,ik}(x,x_1)
   \\ & \notag \times
   \int_{\mathcal{C}} \mathrm{d}x_2 \,
   \Sigma_{kl}'(x_1,x_2) \,
   G_{lj}(x_2,x')
   \>,
\end{align}
and
\begin{align}
   K_i(x,x')
   = - & \int_{\mathcal{C}} \mathrm{d}x_1 \,
         \bigl [ D_0^{-1} + \Pi \bigr ]^{-1}\! (x,x_1) \,
   \\ \notag & \times
         \int_{\mathcal{C}} \mathrm{d}x_2 \,
         \bigl [ \bar K_{0\,k}^{-1} + \Omega_k \bigr ] (x_1,x_2) \,
         G_{ki}(x_2,x')
   \>,
\end{align}
with the modified polarization
\begin{align}
   &
   \Pi'(x,x')
   =
   \Pi(x,x')
   - \int_{\mathcal{C}} \mathrm{d}x_1 \,
      \bigl [ \bar{K}_{0\,k}^{-1} + \Omega_k \bigr ](x,x_1)
   \\ \notag & \times
      \int_{\mathcal{C}} \mathrm{d}x_2 \,
      \bigl [ G_{0\,kl}^{-1} + \Sigma_{kl} \bigr ]^{-1}\!(x_1,x_2)
      \,
      \bigl [ K_{0\,l}^{-1} + \bar{\Omega}_l \bigr ](x_2,x')
   \>,
\end{align}
and self-energies
\begin{align}
   \label{e:Sigmap}
   &
   \Sigma_{ik}'(x,x')
   =
   \Sigma_{ik}(x,x')
    - \int_{\mathcal{C}} \mathrm{d}x_1 \,
      \bigl [ K_{0\,i}^{-1} + \bar{\Omega}_i \bigr ](x,x_1)
   \\ \notag & \times
      \int_{\mathcal{C}} \mathrm{d}x_2 \,
      \bigl [ D_0^{-1} + \Pi \bigr ]^{-1}\!(x_1,x_2) \,
      \bigl [ \bar{K}_{0\,k}^{-1} + \Omega_k \bigr ](x_2,x')
   \>.
\end{align}
These update equations must be solved in conjunction with the
one-point functions, Eqs.~\eqref{e:GammachieqBVA}.  In the
approach of Ref.~\cite{r:AABBS}, the same equations apply with the
restriction that $\Omega_i = 0$ and in the equation for
$\Sigma_{ik}'$ in Eqs.~(\ref{e:SigmasBVA},\ref{e:Sigmap}), the
$\bar{K}_i K_k$ term is absent.

%
%
\section{Numerical results and conclusions}

We ran simulations for a single component (N=1) $\lambda \, \phi^4$
classical field theory, comparing our results to exact Monte Carlo
classical calculations, as described in Ref.~\cite{r:MD02}. The
classical equations become parameter free with a rescaling of
variables as discussed in Ref.~\cite{r:ABW01}.  In particular $x,t
\rightarrow \mu x,\mu t$ and $\phi \rightarrow \sqrt{3 g/\mu^2} \phi$.
Thus by choosing $g = 1/3$, then $\phi$ and time are measured in units
of $\mu$ and $1/\mu$.  For our initial conditions, we took a thermal
gaussian density matrix centered about $\phi(x) = \phi_0$ and
$\dot{\phi}(x) = \pi_0$, thus we are only considering spatially
homogeneous evolutions of $\phi$.  The details of how to initialize
the Green function equations, as well as the solutions for the
symmetric case, are discussed in Ref.~\cite{r:BCDM01}. The numerical
procedure for solving the BVA equations is described in
Refs.~\cite{r:MM02,r:MS02}.  Our update procedure conserved energy to
better than five significant figures.

The large scale calculations presented in this paper, where the
full set of BVA equations is numerically solved, are extremely
cost intensive on a computer. Hardware constraints impose an upper
limit to the extent to which we can follow the time evolution of
the various expectation values. In the following, we choose to
describe the time evolution of the system until we observe the
thermalization of  $\langle \phi^2(t) \rangle$ in the case of the
exact lattice calculation. Since in the quantum case we expect
thermalization to occur faster than in the classical field theory
case, we believe that this criteria sets a useful time scale
for future calculations in the case of realistic quantum field
theory models. In other words, this is the minimum time interval
for which we must be able to follow the time evolution, in order
to make useful predictions in the quantum case \cite{r:CDMq02}.  

\begin{figure}[t]
   \centering
   \includegraphics[width=3.3in]{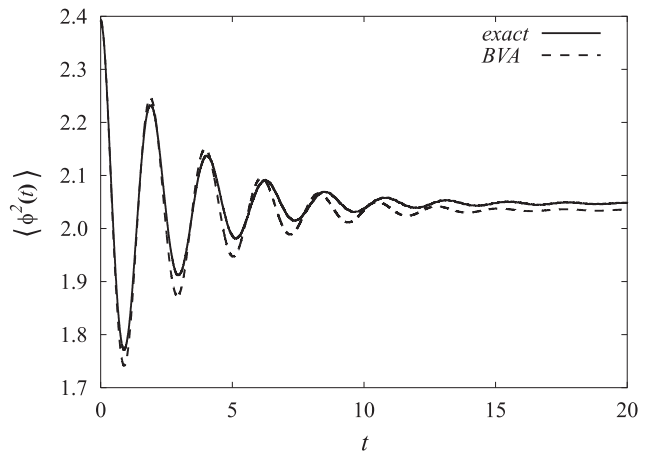}
   \caption{Results for the symmetric case, $\phi(0) = \dot{\phi}(0) =0$.}
   \label{f:fig2}
\end{figure}

In the symmetric case, the two approximations are very similar and
lead to the results shown in Fig.~\ref{f:fig2}.  We notice that
the time-evolution of the two-point function is well described by
the BVA.

\begin{figure}[t]
   \centering
   \includegraphics[width=3.3in]{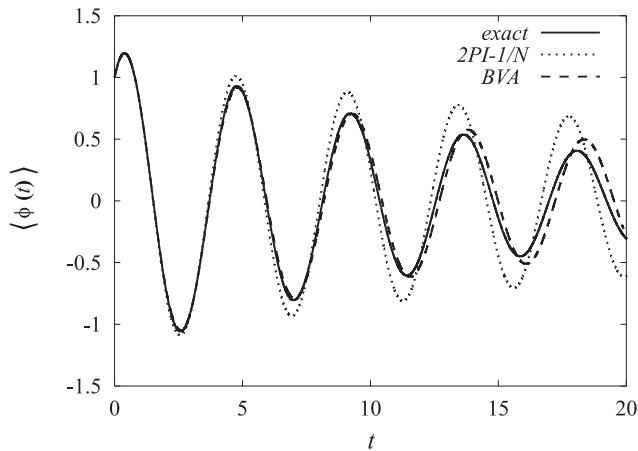}
   \caption{Plot of $\langle \phi(t) \rangle$ for $\phi(0) = 1$, and
   $\dot{\phi}(0) = 1$.}
   \label{f:fig3}
\end{figure}

For the non-symmetric case, we find that the time evolution of the
order parameter $\phi(t)$ is better described by the BVA. For the time
duration required by the exact calculation for the two-point function
to thermalize, the quality of the agreement is similar for all initial
values of $\phi(t)$. Typical results are shown in Fig.~\ref{f:fig3}.
Here we plot the time evolution of $\langle \phi(t) \rangle$ for a
case when the initial symmetry breaking is particularly big.  Despite
the fact that both the BVA and 2PI-1/N expansion fail to accurately
describe $ \langle \phi^2(t) \rangle$, the BVA reproduces the time
evolution of~$\langle \phi(t) \rangle$ quite well for almost the
entire interval of interest.  

\begin{figure}[!h]
   \centering
   \includegraphics[width=3.3in]{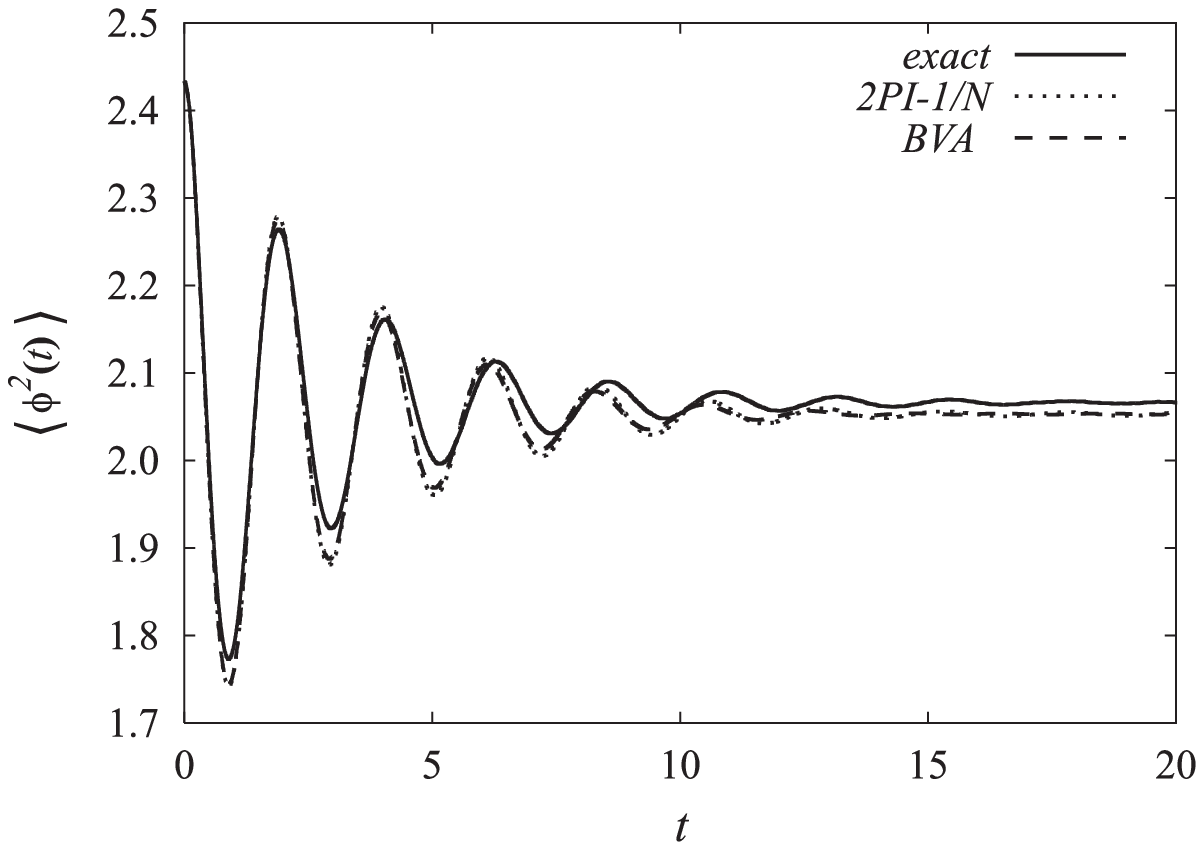}
   \caption{Plot of $\langle \phi^2(t) \rangle$ for $\phi(0) = 0.2$, and
   $\dot{\phi}(0) = 0$.}
   \label{f:fig4}
\end{figure}

\begin{figure}[!h]
   \centering
   \includegraphics[width=3.3in]{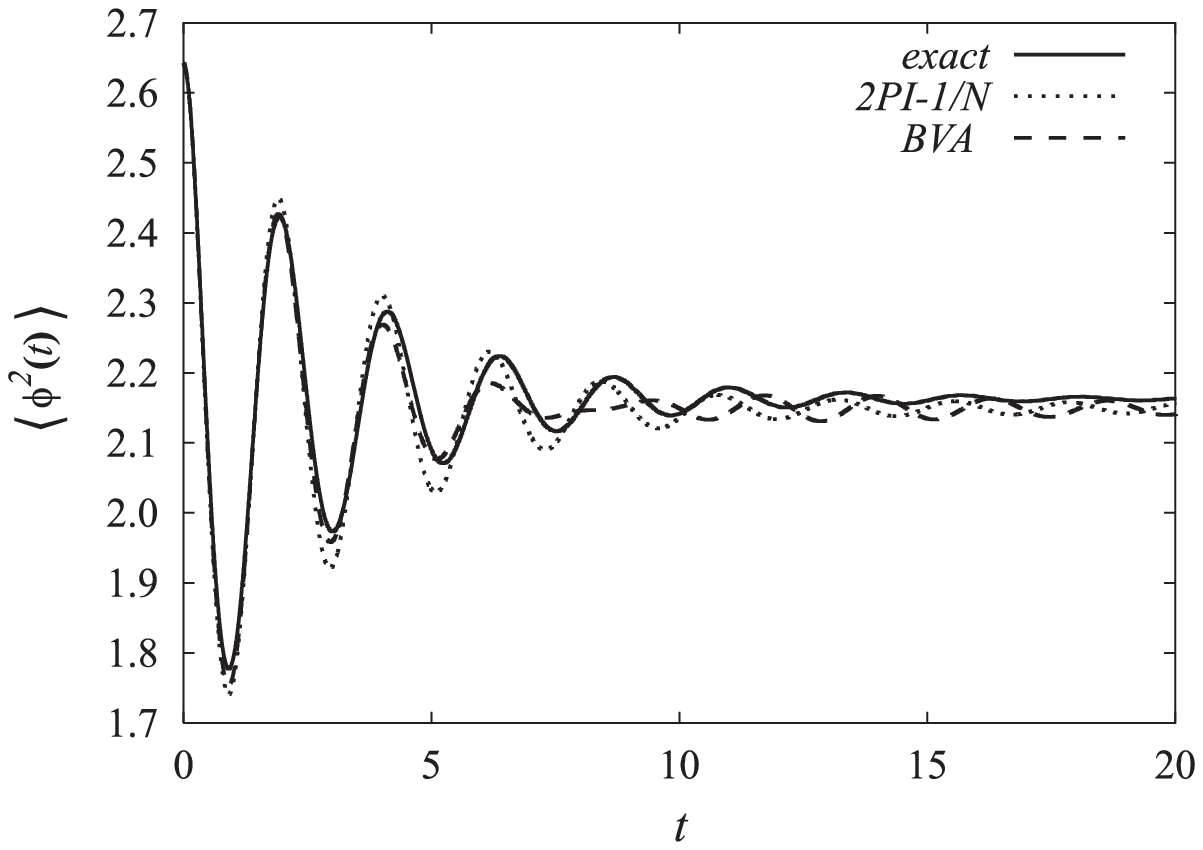}
   \caption{Plot of $\langle \phi^2(t) \rangle$ for $\phi(0) = 0.5$, and
   $\dot{\phi}(0) = 0$.}
   \label{f:fig5}
\end{figure}

\begin{figure}[!h]
   \centering
   \includegraphics[width=3.3in]{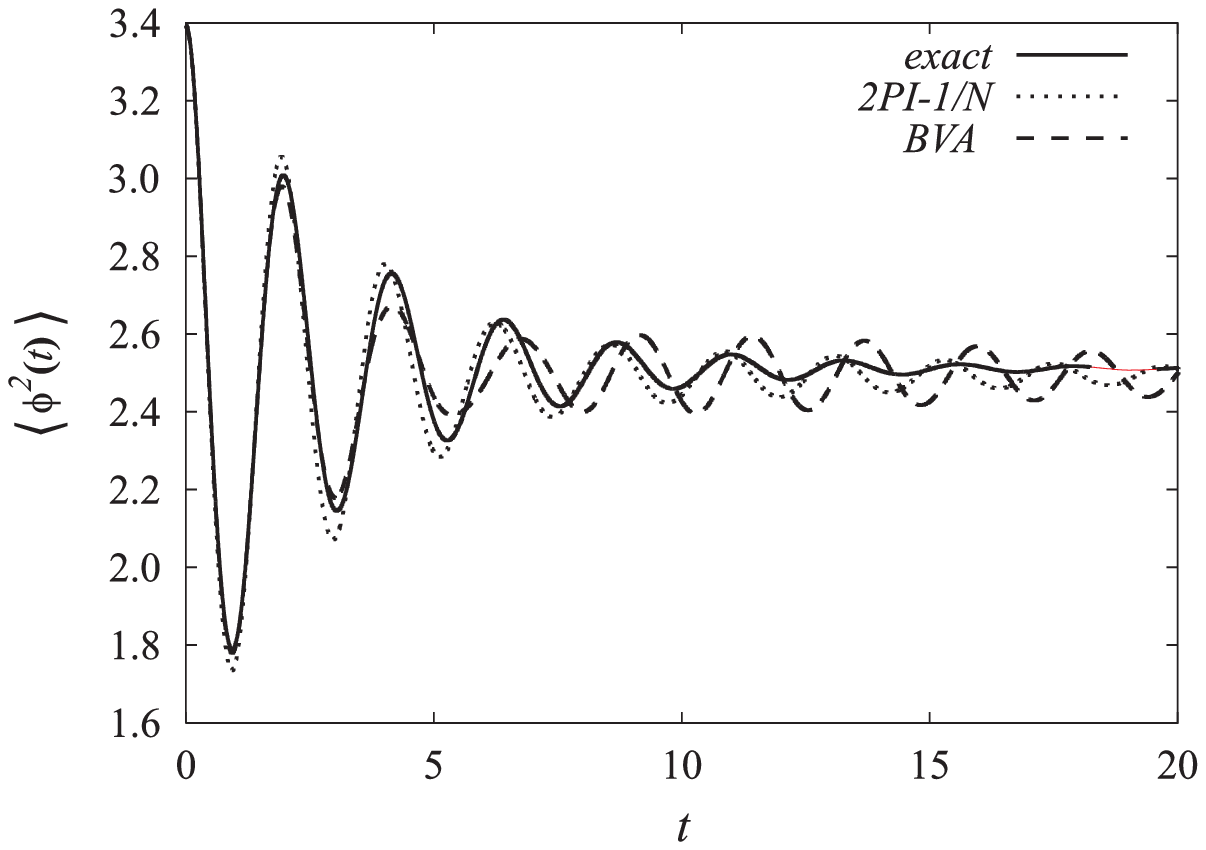}
   \caption{Plot of $\langle \phi^2(t) \rangle$ for $\phi(0) = 1$, and
   $\dot{\phi}(0) = 0$.}
   \label{f:fig6}
\end{figure}

For the two-point correlation function the situation is more
complicated and depends on the size of the initial symmetry breaking
order parameter. The 2PI-1/N expansion thermalizes for all sets of
initial conditions we studied, however the time scale for the
equilibration process of the two-point function, is not reproduced at
large initial symmetry breaking values $\phi(0) \gtrsim 0.4$.  The BVA
gives an equally good agreement as the 2PI-1/N expansion at small
values of $\phi(0)$.  However at larger values of $\phi(0)$, the BVA
exhibits a non-exponentially decaying oscillatory behavior at
intermediate times, before thermalization occurs.  This behavior is
illustrated in Figs.~\ref{f:fig4}--\ref{f:fig6}, where we gradually
increase the initial value of $\phi(0)$.  For simplicity, we take
$\dot{\phi}(0) = 0$ for these cases.
We see in these figures that the 2PI-1/N result is qualitatively better
than the BVA.  However, it is important to note that for large values
of $\phi(0)$, \emph{neither} approximation gives the correct damping
behavior of the two-point function.  This is due to the fact that
three-point vertex corrections become important at intermediate time
scales.  Thus we conclude that although the BVA determines quite well
the time evolution of one-point function (the order parameter here),
the time evolution of the two-point function requires a better
treatment of the three-point vertex.

The accuracy of the BVA depends on the size of the initial symmetry
symmetry breaking.  This seems to indicate that for a regime where
thermalization occurs relatively slowly, such as in the case of this
classical field theory model, higher order corrections do eventually
become increasingly important at later times.  The quantum field
theory case is qualitatively different.  In the quantum case,
thermalization is expected to settle in much faster.  Therefore, one
can hope that in the quantum case the noise will not have enough time
to buildup and consequently will result in a lesser distortion of the
BVA predictions.  Future studies must go beyond the BVA and
investigate the magnitude of these further corrections, especially in
the quantum case.

In the above approximations, we set the vertex function to its
bare value of unity.  However the formalism tells us how to
determine the corrections contained in~$\Gamma_2$.  The
three-point vertex function is found from the
self-energy~$\Sigma_{\alpha\beta}$ by:
\begin{multline}
   \Gamma_{\alpha\beta\gamma}(x,x',x'') \\
   =
   f_{\alpha\beta\gamma} \, \delta(x,x') \, \delta(x,x'')
   +
   \frac{ \delta \, \Sigma_{\alpha\beta}(x,x') }
        { \delta \phi_{\gamma}(x'') } \>,
   \label{e:vert}
\end{multline}
where $f_{0,i,j} = f_{i,0,j} = f_{i,j,0} = \delta_{ij}$.  The
integral equation generated from this equation for
$\Gamma_{\alpha\beta\gamma}(x,x',x'')$ is discussed in
Ref.~\cite{r:MCD01}.  Thus one way of including vertex corrections
is to use the Green functions calculated here to obtain an
approximate nonlocal vertex function from Eq.~\eqref{e:vert},
reintroduce the vertex function into the exact SD equation for the
two-point function equation and iterate until a stable solution is
found.  A second approach is to include three loop terms in
$\Gamma_2$ and solve the resulting set of equations.  These
approaches will be explored elsewhere.

To conclude, in our quest for an accurate assessment of NLO versus
mean-field effects in quantum field theory, the present work
represents an important stepping stone. In this paper we report
the first results of the full BVA calculation, and in doing so we
conclude work done earlier in a more restrictive
regime~\cite{r:BCDM01}.  Our goal remains the study of the quantum
field theory case, but the classical field theory case is still
interesting since this is the high temperature limit of the
quantum case. We want to make sure that we know to what extent
this limit is reproduced. This is even more important since this
is the only case for which exact calculations are possible, and
thus the classical field theory O(1) model with one spatial
dimension represents a key benchmark.

Having established the domain of applicability of the BVA in this
extreme limit, it is clear that two avenues must be explored in the
near future: first we will apply the BVA formalism to the study of the
quantum linear sigma model, and other realistic models, in 3+1
dimensions, and in doing so we will obtain first insights into the
reliability of mean-field approaches in studies of the real-time
evolution of realistic quantum field theory models.  Next we will
investigate the role of further corrections to the BVA, which will be
the real test of the accuracy of the BVA formalism.  This project will
involve the development of practical numerical algorithms and,
possibly, require the next generation of supercomputers.

%
%
\begin{acknowledgments}

Numerical calculations are made possible by grants of time on the
parallel computers of the Mathematics and Computer Science
Division, Argonne National Laboratory. The work of BM was
supported by the U.S. Department of Energy, Nuclear Physics
Division, under contract No. W-31-109-ENG-38.  JFD and BM would
like to thank Los Alamos National Laboratory and the Santa Fe
Institute for hospitality.

\end{acknowledgments}

\vfill
%
%
%
%
\bibliography{johns}
%
%
%
\end{document}